# Ultra-Fast and Efficient Design Method Using Deep Learning for Capacitive Coupling WPT System

Rasool Keshavarz, Ehsan Majidi, Ali Raza, and Negin Shariati

*Abstract*— **Capacitive coupling wireless power transfer (CCWPT) is one of the pervasive methods to transfer power in the reactive near-field zone. In this paper, a flexible design methodology based on Binary Particle Swarm Optimization (BPSO) algorithm is proposed for a pixelated microstrip structure. The pixel configuration of each parallel plate (43×43 pixels) determines the frequency response of the system (S-parameters) and by changing this configuration, we can achieve the dedicated operating frequency (resonance frequency) and its related |S$_{21}$| value. Due to the large number of pixels, iterative optimization algorithm (BPSO) is the solution for designing a CCWPT system. However, the output of each iteration should be simulated in electromagnetic simulators (e.g., CST, HFSS, etc.), hence, the whole optimization process is time-consuming. This paper develops a rapid, agile and efficient method for designing two parallel pixelated microstrip plates of a CCWPT system based on deep neural networks. In the proposed method, CST-based BPSO algorithm is replaced with an AI-based method using ResNet-18. Advantages of the AI-based iterative method are automatic design process, more efficient, less time-consuming, less computational resource-consuming and less background EM knowledge requirements compared to the conventional techniques. Finally, the prototype of the proposed simulated structure is fabricated and measured. The simulation and measurement results validate the design procedure accuracy, using AI-based BPSO algorithm. The MAE (Mean Absolute Error) of prediction for the main resonance frequency and related |S$_{21}$| are 110 *MHz* and 0.18 *dB*, respectively and according to the simulation results, the whole design process is 3629 times faster than the CST-based BPSO algorithm.**

## I. INTRODUCTION

THE use of smart systems including cell phones, remote devices, and wireless sensors is increasing rapidly every year [1]. All these wireless devices are powered using short-life batteries which need to be recharged or replaced after a certain amount of time [2]. For Internet-of-Things (IoT) wireless sensor networks (WSNs), it is impossible to replace batteries in some applications where accessibility is challenging, such as implantable devices including a heartbeat pacemaker [3]. Wireless power transfer (WPT) is a feasible solution to avoid battery problems through powering devices wirelessly[4], [5].

WPT systems can be divided into two main streams i.e., far-field, and near-field including radiating and non-radiating (reactive) near-field WPT [6]. In far-field and radiating near-field WPT systems, the distance between transmitter and receiver antennas (*d*) with size of D should be $d > 2D^2/\lambda$ and $0.62\sqrt{D^3/\lambda} < d < 2D^2/\lambda$ where $\lambda$ is the wavelength. Although in reactive near-field WPT, transmitter and receiver plates or coils should be placed as d<$0.62\sqrt{D^3/\lambda}$ [7]. Due to the short distance between Tx and Rx in the near-field WPT systems, the conversion efficiency is high (more than 50%). Therefore, reactive near-field WPT has been used in biomedical applications such as body implants for detection and treatment of diseases [8-11].

Regarding the coupling mechanism between Tx and Rx sections, reactive near-field method divides into two methods as inductive or capacitive coupling methods [12, 13]. In inductive coupling WPT (ICWPT), power is transmitted from source to receiver through coils using a magnetic field. In this scenario, high transmission power can be achieved when both coils are close to each other. In capacitive coupling WPT (CCWPT), power is transferred through electric fields using two parallel metal plates like electrodes, which are placed in the reactive near-field zone of each other [14, 15]. CCWPT has some special privileges over ICWPT as the electric field is largely confined between the capacitor plates. Hence, reducing interference and providing higher immunity for misalignment issues between the transmitter and receiver [16, 17]. Whereas, ICWPT efficiency strongly depends on the alignment of transmitter and receiver coils [18].

Capacitive coupling is generally done at MHz frequency ranges [19]. Recently, RF/microwave WPT systems are proposed for low power IoT devices and implants. Specifically, when the Tx/Rx are limited to millimeter-size, high frequency WPT systems are inevitable [20-24]. Moreover, radiating near-field and far-field RF/microwave WPT systems are subjected to the constraints of high propagation loss (path loss) between Tx

RF and Communication Technologies (RFCT) research laboratory, University of Technology Sydney, Ultimo, NSW 2007, Australia, e-mail: Rasool.Keshavarz@uts.edu.au.

and Rx sections [25-27]. Therefore, RF/microwave reactive near-field WPT system is a good candidate to realize a compact and highly efficient WPT scenario. Further, high frequency WPT systems are able to be used for simultaneous information and power transfer (SWIPT) scenarios for low power devices and implants. It can provide additional inherent advantages such as high data rate, less sensitivity in the misalignment between Tx and Rx plates/antennas, and good isolation between the power transfer and communication circuitry [28,29].

Wireless Power Transfer (WPT) using couplers enables energy transmission between a source and receiver without physical connections, making it ideal for applications where wired connections are impractical. This technique, operating within a few millimeters to centimeters, is employed in wireless charging for devices like smartphones, while also promising safer and less invasive power supply for implantable medical devices like pacemakers. The proposed system focuses on short-distance WPT, with coupling effectiveness decreasing as plate distance increases.

Conventionally, the size of broadside coupling structures in CCWPT mainly depends on the operating frequency (inverse relationship) [30]. However, the size of the structures is pre-defined in some applications such as implantable devices where size is the main constraint. In such applications, the designing of CCWPT cannot be done using conventional methods where sizes are frequency dependent. Hence, a new flexible approach is required to design the WPT system.

Moreover, conventional designing of different RF structures (antennas, metasurfaces, RF filters, couplers, etc.) is a time-consuming process involving modeling, simulation, and optimization. This also needs a strong background knowledge of EM (electromagnetics). A parametric sweep method allows for a parameter to be swept through a range of values and is useful for finding the optimum value and sensitivity of the design performance to certain parameters. However, parametric sweep is a time-consuming process and is not a practical solution for an optimization problem with multiple input parameters. Therefore, to simplify the design process, iterative optimization algorithms have been used in RF (radio frequency) structure and circuit design. Most common optimization algorithms are Genetic Algorithms, Firefly Algorithm and Particle Swarm Optimization (PSO) [31,32]. Optimization algorithm is applied by running a large number of simulations with an advanced optimization algorithm.

Recently, pixelated microstrip structures have attracted attention of EM researchers due to its effectiveness and flexibility in designing different types of microwave devices [31,32]. The pixelation means breaking down a microstrip area into small size rectangle pieces. The concept of pixelating an area within the structure is to define binary numbers "1" and "0" which refer to metal and no metal rectangles, respectively. In the proposed structure, binary PSO (BPSO) helps us to find the optimum pixel configuration to reach the predefined goal (maximum |$S_{21}$| at a specific frequency band).

The choice of Binary Particle Swarm Optimization (BPSO) was motivated by its specific advantages in dealing with binary search spaces. While other algorithms like Genetic Algorithms (GA) can also handle binary spaces, BPSO was preferred due to its inherent ability to explore and exploit the solution space efficiently, particularly in our context of optimizing pixel configurations for wireless power transfer systems. BPSO's fast convergence, simplicity in implementation, and potential for finding optimal solutions within a constrained binary domain played a significant role in its selection [31, 32].

In microstrip design, the BPSO algorithm follows these steps: It initiates with the creation of a population of binary-encoded particles, each representing a potential solution. These particles are initialized randomly with binary values, corresponding to design parameters like microstrip pixel dimensions. The fitness of each particle is evaluated based on an objective function that assesses microstrip performance metrics. Particle positions are updated iteratively, influenced by inertia, cognitive (individual best) and social (global best) components, driving the search for improved solutions. Given its focus on binary optimization, particle positions are represented as binary strings. The algorithm repeats its iterations until a convergence criterion is met or a predetermined number of iterations is reached. The termination leads to a solution where the best particle position reflects the optimized microstrip design.

Iterative optimization algorithms like BPSO run several iterations to achieve predefined goals. Then, the extracted configuration of pixelated microstrip structure should be simulated in electromagnetic (EM) simulators (CST, HFSS and ADS) to find S-parameters in each iteration. The results are then returned to the optimization algorithm to be used in the next iteration. Since running EM simulation for each iteration is also a time-consuming process, finding an efficient solution to speed up the simulation process of each iteration is tangible.

In this work, capacitive coupling plates are considered as pixelated microstrip structures to achieve flexibility in the design procedure in terms of operating frequency, size and coupling level. Proposed pixelated structure size of $62 \times 62\ mm^2$ is divided into 43×43 pixels in a symmetrical pattern as 4 quadrants and 36 tiles. The proposed structure can achieve pixelated broadside coupler in a specific frequency range within 1 to 5 $GHz$ with different coupling levels.

To reduce the computational cost of iterative algorithm and speed up EM simulations, Convolutional Neural Networks (CNNs) are used to predict the S-Parameters of broadside coupler. Several machine learning techniques have been reported for EM structure design [33-36], but very less work has been focused on the optimization of EM structures using adaptive machine learning [37]. Specifically, the CNN approach has not been used previously on pixelated EM structures for CCWPT applications.

In this work, the proposed architecture comprises of ResNet-18 to extract feature maps followed by two fully connected layers [38, 39]. We construct reference datasets of 1000 samples from randomly generated symmetrical pixels using CST software. Therefore, a rapid, efficient and flexible design method is proposed for CCWPT system based on pixelated microstrip plates, while CST simulator is replaced with an AI-based method. In the proposed pixelated CCWPT system, size, frequency, and coupling level can be considered as the input parameters for the AI-based method as shown in Fig. 1.

Major contributions of this paper are summarized as follows:
- For the first time, the pixelization technique using BPSO optimization method is proposed to implement a near-field CCWPT system. In the proposed technique, CST software is replaced with AI algorithm to return the |S|-parameter of the pixelated structure to the iterative algorithm (BPSO) in each iteration.
- The proposed optimization method based on deep learning, substitutes CST simulation in the design process of pixelated CWPT using BPSO algorithm.
- Efficient, agile and rapid design methodology for pixelated microstrip structures is proposed based on deep learning.
- The proposed technique enables quick and accurate design of the CCWPT system to achieve desired efficiency and operating frequency within a predefined dimension.
- Even though in this paper the technique is applied only to a CCWPT system as an application, the flexible design methodology based on pixelated microstrip structure and ResNet-18 network can be adapted to different structures. This enables us to precisely develop other RF and microwave structures (e.g. filters, duplexers, antennas, metasurfaces, etc.) at different frequency bands and desired dimensions.

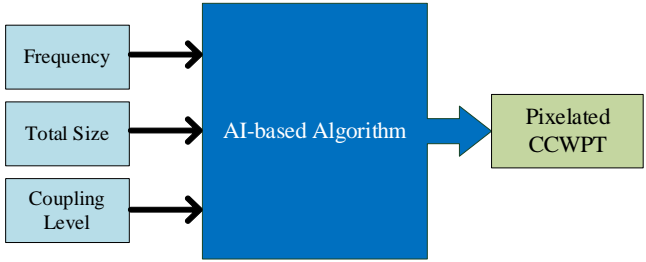

Fig. 1. Design approach of pixelated CCWPT system using AI algorithm

The organization of this paper is as follows: CCWPT system design process and methodology based on deep learning are presented in Section II. The proposed CCWPT system performance is validated by simulation and measurement in Section III. Further, the proposed AI-based method results are provided in this section. Lastly, the conclusion is presented in Section IV.

## II. PROPOSED CCWPT SYSTEM: THEORY AND DESIGN USING RESNET

Fig. 2 shows the schematic and equivalent circuit model of the proposed CCWPT system which includes two identical pixelated microstrip plates. A specific pixel pattern determines the operating frequency and efficiency of the CCWPT system for a fixed patch area ($L \times W$) and distance ($d$). The output port of the receiver is connected to a rectifier circuit to convert RF signal to DC voltage [40]. An impedance matching circuit is used to ensure maximum power transfer between the receiver and rectifying circuit as shown in Fig. 2(a). Proposed pixelated microstrip plates are modeled as parallel $LC$ tank circuit ($L_P$ and $C_P$) and $C_g$ is the capacitance between two parallel patches as shown in Fig. 2(b). $C1$ and $C2$ are the capacitances of voltage doubler and rectifier sections, respectively.

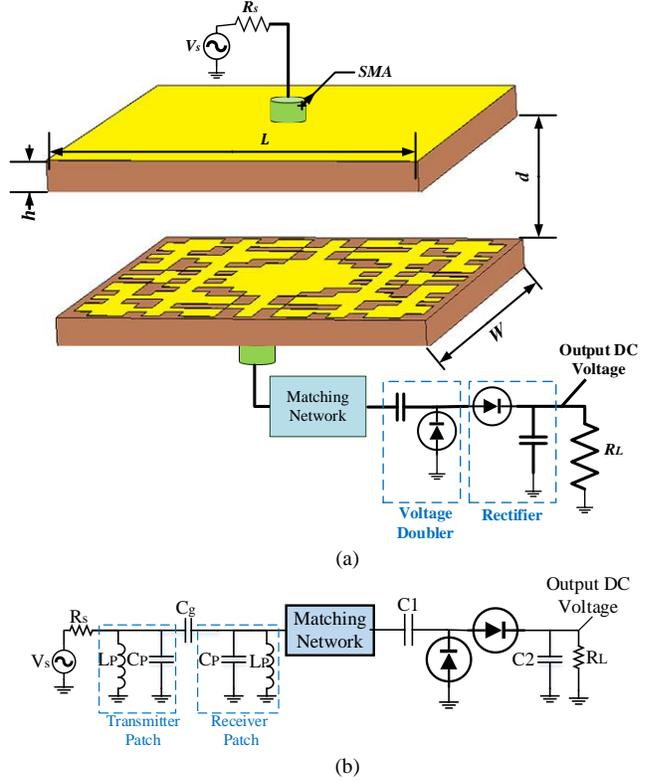

Fig. 2. Proposed CCWPT system, a) 3D view, b) equivalent circuit model.

### A. Structure of the Proposed Pixelated Microstrip Plates

Fig. 3 shows the pixel configuration of the proposed pixelated microstrip plates. To achieve insensitivity to misalignment between the positions of plates, the top layer of each microstrip patch is composed of 4 quadrants which are mirrored about the *x* and *y*-axis for a symmetrical pattern. Each quadrant consists of three random tiles ($T_1$, $T_2$ and $T_3$). The tiles are placed in a symmetrical configuration as shown in Fig. 3. There are a number of constrains that should be considered to select a proper pixel configuration, number of pixels, and total dimension:
- To achieve a homogenous structure based on the electromagnetic theory [41], the pixel size ($a_p$) should be very smaller than wavelength at maximum operating frequency range ($a_P \ll \lambda_g \rightarrow a_P < \lambda_g/20$).
- Maximum frequency in the design procedure is 5 GHz and FR-4 permittivity is 4.3, hence, $\lambda_g (at\ 5\ GHz) = 28\ mm$ and the pixel size should be less than 2.8 mm, where we chose 1.4 mm.
- The proposed pixelated WPT system needs to cover a frequency range of 1 to 5 GHz. Therefore, the total structure size should be capable to resonate at all frequencies in this range. The dominant frequency for the size of whole structure is the lowest frequency band, 1 GHz. The wavelength at 1 GHz equals 142 mm on FR-4 substrate and for a simple microstrip structure the resonance frequency occurs for patch size around $\lambda_g/2$. Therefore, the whole size of the pixelated structure to be capable of resonating at 1 GHz on FR-4 substrate is around 72 mm. We selected the total size 62 mm to show the compactness of the proposed structure compared with a simple patch structure.

- By increasing the number of pixels, the total considerable cases increase, and so the training process of AI algorithm will become a time-consuming procedure. Therefore, we considered 43×43 pixels on a 62×62 mm plate.
- To overcome the misalignment problem in a WPT system, a symmetry structure is considered for pixelated plates. Hence, the plate is divided into four quadrants and all quadrants are the same.
- Each quadrant is constituted from 3×3 tiles, to reduce the number of cases in the training process of AI algorithm and saving time.
- Finally, to conduct input power to all quadrants, the "+" section at the center of structure is fixed for all pixel configurations.

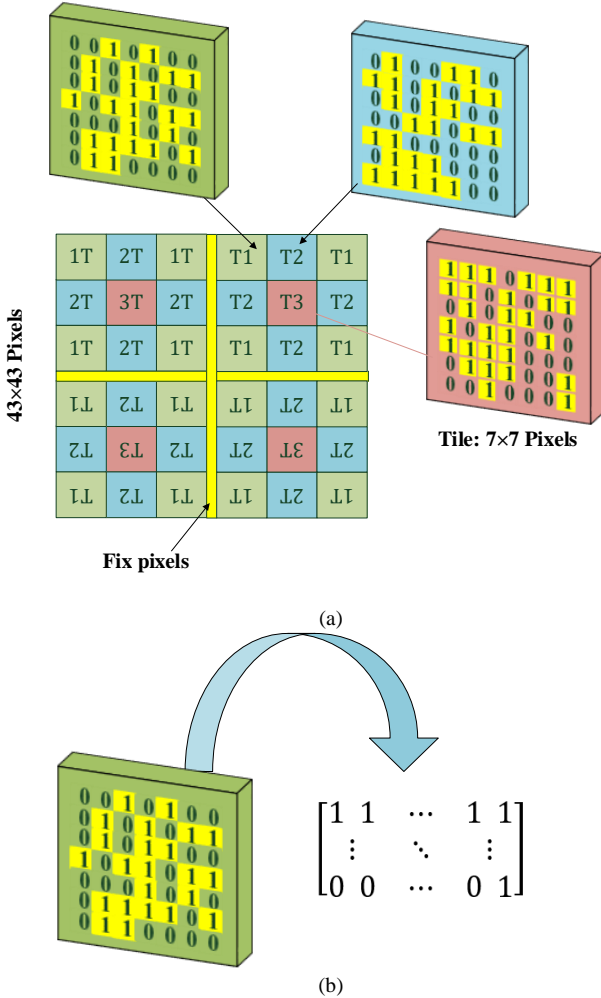

Fig. 3. Pixel configuration of the proposed pixelated microstrip plates, a) placement of three different tiles at the top layer of the symmetrical patch, b) pixel configuration of each tile and matrix representation.

Therefore, each tile has a total of 7×7 pixels where "1" means the area is covered by copper, whereas "0" pixel is no copper area as shown in Fig. 3 (b). The quadrants are separated using fixed pixels of "1s" for all configurations. In this way, the unit cell structure can be encoded by a 43×43 matrix which is used in the proposed algorithm.

Since there are 49 pixels in each tile, the number of possible tiles is $2^{49}$. Due to three random tiles in a quadrant ($T_1$, $T_2$, and $T_3$), the total number of considerable structures is $1.7841 \times 10^{44}$ which is fairly huge. This would take thousand trillion of years to cover all the data to design a pixelated structure using conventional methods which is an impossible task with available resources. Besides, it is difficult to directly discover the rules between pixelated microstrip plate matrices and electromagnetic properties of the structure (S-parameters). Therefore, deep learning can discover the rules between the pixelated microstrip plate and electromagnetic properties and predict the S-parameters of each parallel pixelated microstrip plates.

*B. Data Collection*

For the training process, 1000 configurations of pixelated structures (43×43) are randomly generated using MATLAB software and the S-parameters of the parallel pixelated microstrip plates are simulated in EM simulation software CST Microwave Studio (MWS). The captured data is used to train a deep neural network. The volume of data is determined through theoretical analysis which demonstrates the adequacy of 1000 sets of samples as training data.

Fig. 4 shows the probability density function (PDF) and cumulative density function (CDF) for 1000 different configurations of pixelated microstrip plates. For extracting the proper PDF and CDF distributions, the global maximum point of $|S_{21}|$ (main resonance frequency) for each configuration of the pixelated microstrip plates are considered as samples in the data analysis. Fig. 4(a) and 6(b) exhibit the PDF of the main resonance frequency and associated $|S_{21}|$ values, respectively. Fig. 4(a) shows that the probability distribution of the main resonance frequency is almost uniform which is ideal for our application. According to Fig. 4(b), the $|S_{21}|$ distribution follows a negatively skewed distribution. Skewness is a measure of the asymmetry of the data around the sample mean and can be a positive or negative value. The Gaussian distribution and skewness are defined as:

$$f(x) = \frac{1}{\sigma\sqrt{2\pi}} \exp\left(-\frac{(x-\mu)^2}{2\sigma^2}\right) \quad (1)$$

and

$$s = \frac{\frac{1}{n}\sum_{i=1}^{n}(x_i-\mu)^3}{\left(\sqrt{\frac{1}{n}\sum_{i=1}^{n}(x_i-\mu)^2}\right)^3} \quad (2)$$

where $\mu$ and $\sigma$ are the mean and standard deviation of $x$, respectively. Table I illustrates the mean, standard deviation, skew factor, median and mode of samples for 43×43 pixel configurations.

Fig. 5 shows the joint probability of samples for main resonance frequency and related $|S_{21}|$ value. According to this figure, there is a gap from 3.4 to 3.7 *GHz* where the concentration of samples is very low. Therefore, we restricted the frequency range from 1 to 3.4 *GHz* and 3.7 to 5 *GHz* for training the proposed AI-based algorithm.

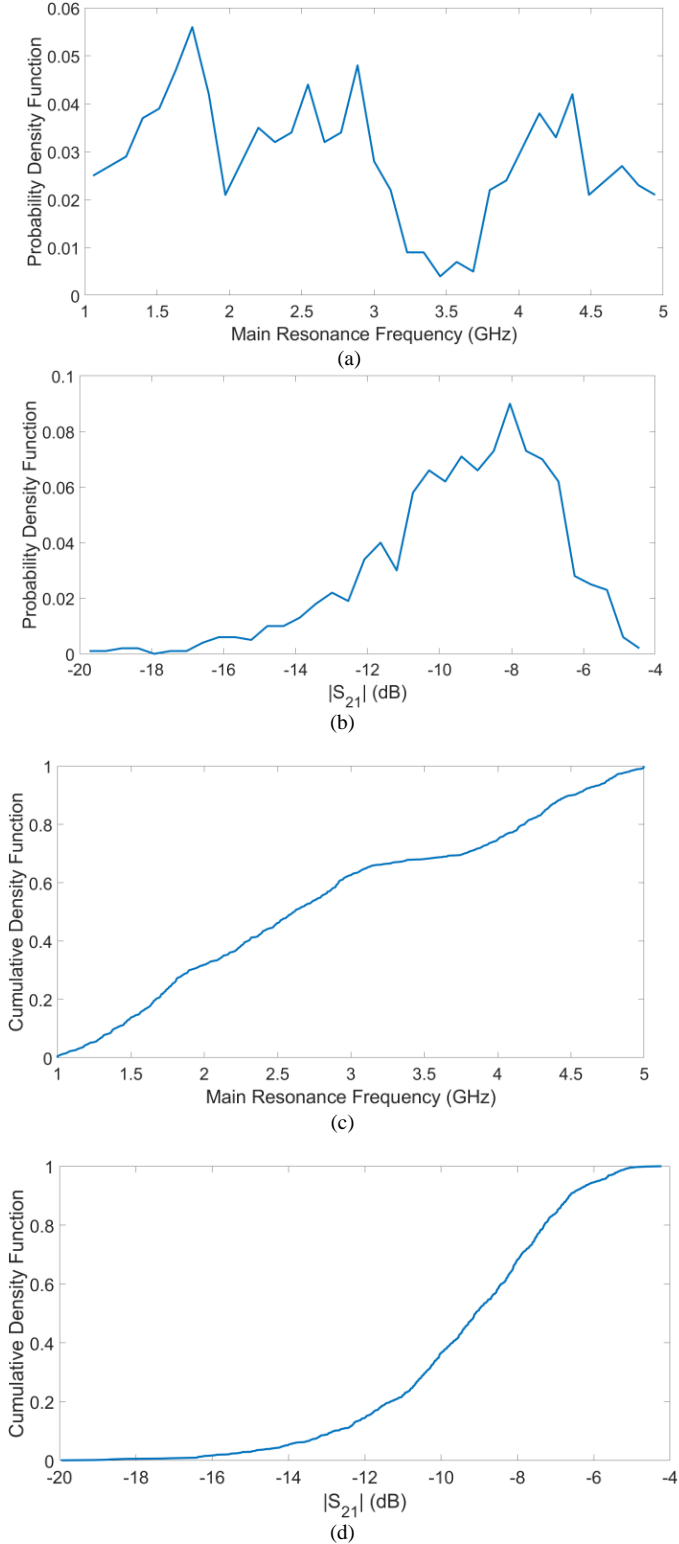

Fig. 4. Probability density functions (PDFs) of a) frequency, b) $|S_{21}|$, cumulative density functions (CDF) of c) frequency, d) $|S_{21}|$ samples.

TABLE I
STATISTICAL PARAMETERS OF SAMPLES FOR 43×43 PIXEL CONFIGURATIONS (1000 DATA)

| Targets | Mean ($\mu$) | Standard Deviation ($\sigma$) | Skew Factor ($s$) | Median | Mode |
|---|---|---|---|---|---|
| Main resonance frequency (*GHz*) | 2.82 | 1.17 | 0.3 | 2.62 | 1.74 |
| $|S_{21}|$ (*dB*) | -9.4 | 2.5 | -0.83 | -9.1 | -8 |

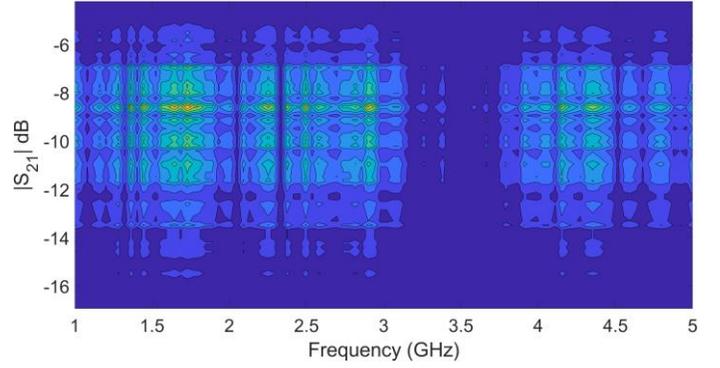

Fig. 5. Joint probability distribution of the main resonance frequency and related $|S_{21}|$.

## C. AI-Based Optimization Method

Fig. 6 exhibits the process of three different methods for designing pixelated microstrip plates. In the conventional method, we need to extract an electromagnetic circuit model for the desired structure which could be complicated for pixelated plates. In other words, finding the values of equivalent circuit model components shown in Fig. 2(b) or applying a distributed model on pixelated structures is difficult.

Therefore, we should use iterative optimization methods to design the pixelated microstrip plates. In this approach, a parametric sweep is performed to attain the best values. If the best values meet the requirements of design target, the design process ends else, a new pixel's configuration will be generated and the design procedure will restart. Nevertheless, the main challenge of using iterative algorithm for designing a pixelated structure is the simulation time of the pixelated microstrip plates using EM simulators (e.g. CST or HFSS) after each iteration. The simulation time of each pixelated structure in CST takes around 15 minutes and iterative algorithm needs about 1000 iterations for converging. Hence, the total simulation time is 250 hours (10 days and 10 hours) which is impractical in most cases. During training, 1000 randomly generated pixelated structures (43x43) undergo electromagnetic simulation in CST Microwave Studio using MATLAB-generated S-parameters. This dataset trains a deep neural network in 250 hours, enabling AI-based optimization to predict and design pixel configurations across the 1 to 5 GHz frequency range.

Therefore, we propose a deep learning method based on Residual Network (ResNet) to realize a rapid and efficient design algorithm for parallel pixelated microstrip plates in the CCWPT system. In this AI-based method, a deep neural network is used instead of CST software which returns the S-Parameter results after each iteration (Fig. 6).

Each pixelated microstrip plate is divided into 43 columns and 43 rows, with a total of 1849 pixels. Any of the pixels can have a value of "0" or "1" assigned randomly. A value of "1" means that the small area allocated to that pixel is covered with metal, and "0" means a normal cell without metal. By changing the values assigned to the pixels on the parallel patches, the coupling level ($|S_{21}|$) is changed. Fig. 7 illustrates the procedure of the proposed method. The target of the proposed method is to meet predefined goals (e.g., operating frequency, efficiency, and size) with high accuracy and in a short time.

The feature extraction architecture is adopted from the ResNet-18 architecture which has been widely successful across a range of computer vision tasks [38]. It includes an initial stemming stage followed by 4 ResNet stages and a final classification layer [39]. Each ResNet stage consists of two residual blocks and each residual block contains 2 convolutional layers. The residual blocks contain shortcut connections between the input and output of each residual block, allowing for better gradient propagation and stable training. A residual block shown in Fig. 7 is defined as [38]:

$$y = \mathcal{F}(x, \theta) + x \qquad (3)$$

Where $x$ and $y$ are the input and output vectors of the block. The function $\mathcal{F}(x, \theta)$ represents the residual mapping to be learned. The operation $(\mathcal{F} + x)$ is performed by a shortcut connection and element-wise addition.

ResNet-18 yields 512 feature maps. We removed the last fully connected layer of ResNet and replaced it with two fully connected layers with 64 and 2 channels. Also, Rectified Linear Init (ReLU) was used as an activation function for the first fully connected layer. So, the total depth of the network is 19 layers (Fig. 7). These architecture hyperparameters were selected via grid search using training and validation split on the regression tasks. In Fig. 7, FC-64 and FC-2 are fully connected layer with number of channels 2 and 64, respectively.

The network on the two isolated regression tasks is trained, where the output of our architecture is a prediction of the main resonance frequency and $|S_{21}|$, to achieve the optimal peak. The SGD (Stochastic Gradient Descent) optimizer is used across all experiments with learning rate 0.001, momentum 0.9, and batch size 256. Then, we split the data into 60/20/20 train, validation and test splits, train the network on the train split, and use the MAE (Mean Absolute Error) on the validation split to perform model selection as follows:

$$MAE = \frac{1}{n}\sum_{i=1}^{n}|y_i - x_i| \qquad (4)$$

Then, MAE on the test dataset is evaluated and reported. According to evaluation reports of MAE on the test dataset, the proposed AI-based iterative method does not need a time-consuming optimization procedure compared to CST-based iterative method. Further, the proposed method simplifies the design process and improves the evaluation speed, efficiency and effectiveness of the iterative methods.

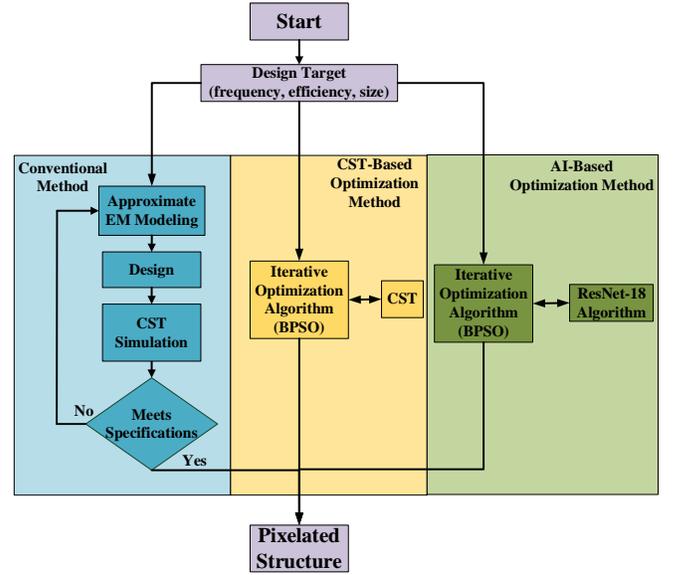

Fig. 6. Different design methods for pixelated structures.

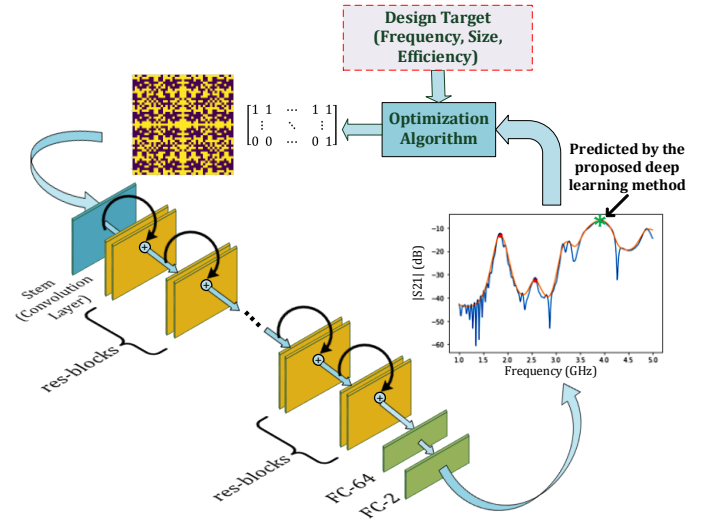

Fig. 7. Working Principle of the proposed deep learning method based on residual network.

### III. SIMULATION, MEASUREMENT AND DISCUSSION

To validate the performance of the proposed CCWPT system and AI-based iterative method for designing parallel pixelated plates, one prototype is fabricated in all cases. Fig. 8 shows the fabricated prototype and measurement setup for the CCWPT system. Microstrip pixelated plates are fabricated on FR4 substrate with thickness $h = 1.6$ mm, dielectric constant $\varepsilon_r = 4.3$, and loss tangent $tan\delta = 0.025$. The thickness of the copper at the top and bottom layers ($T$) is 0.035 mm and the distance between two parallel plates is 5 mm. The aimed specifications of the proposed CCWPT system are resonance frequency around 1.8 GHz, whole plate size of 62 mm × 62 mm, high efficiency at 0 dBm (more than 50%), while the thickness of foam (medium between two plates) is 5 mm. Table II shows the geometry of the fabricated CPWT system of Fig. 8.

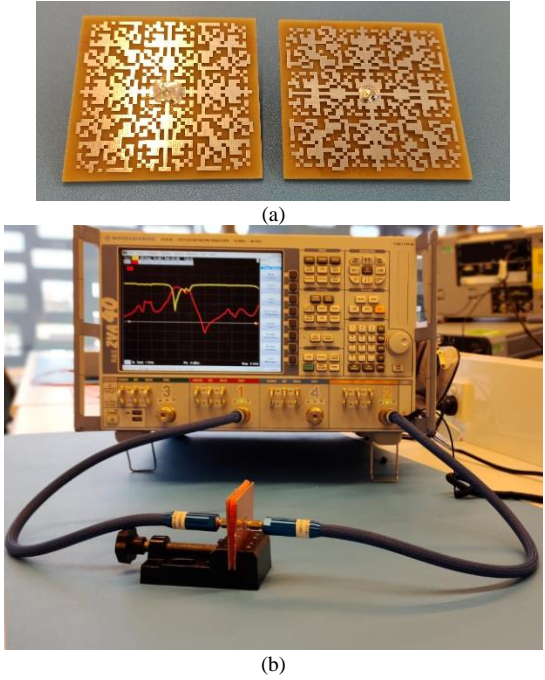
(a)

(b)

Fig. 8. a) Fabricated prototype, b) measurement setup.

TABLE II
DIMENSION OF THE PROPOSED CCWPT (IN MILLIMETER)

| h | L | d | W | T | Pixel size |
|---|---|---|---|---|---|
| 1.6 | 62 | 5 | 62 | 0.035 | 1.4 |

### A. Simulation and Measurement Results of Parallel Pixelated Microstrip Plates

Fig. 9 shows the electric field (E-field) and magnetic field (H-field) distributions of each parallel plane in the proposed CCWPT system and offers valuable insights into the propagation characteristics of electromagnetic waves along the structure. The E-field distribution illustrates the varying intensity of the electric field along the microstrip, highlighting regions of high and low field strength. This aids in understanding how energy is distributed across the line's dielectric substrate and conductive traces. On the other hand, the H-field distribution provides a visual representation of the magnetic field's behavior perpendicular to the electric field, shedding light on the circular loops of magnetic flux surrounding the microstrip conductor. The interaction between the E-field and H-field distributions influences the impedance, signal transmission, and radiation properties of the microstrip line.

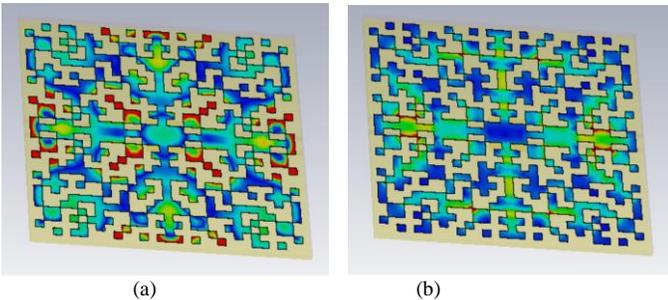
(a) (b)

Fig. 9. E-filed and H-field distribution of each plane at resonance frequency (1.77 GHz).

Simulated and measured S-parameter results of the fabricated prototype are shown in Fig. 10. According to this figure and measurement results, the main resonance frequency of the fabricated prototype is 1.77 *GHz* and the corresponding $|S_{21}|$ is better than -3 *dB*. Also, return loss at operating frequency bands is greater than 19 *dB*. Moreover, a good agreement between simulation and measurement results confirms the consistency of the design and measurement process. In addition, the black star (*) in Fig. 10 exhibits the predicted value of the proposed AI-based method at the main resonance frequency and corresponding $|S_{21}|$. The MAE error of predicted resonance frequency and corresponding $|S_{21}|$ are 110 *MHz* and 0.18 *dB*, respectively (Table III). Hence, there is a good agreement between the proposed AI-based method and design target.

Fig. 2 and Fig. 3 illustrate the adoption of a symmetric quadrant-based approach for pixelated plates to address misalignment in WPT systems, influencing the design process. Therefore, another benefit of the proposed CCWPT system is its low sensitivity to misalignment and rotation of one side with respect to the other plate. Fig. 11 shows the measurement results of misalignment for different α (rotation angle) and β (amount of vertical displacement) values. According to the results, the proposed CCWPT system is almost insensitive to rotating the plates for $α= 0^0, 45^0, 90^0$ and displacement in x and y directions ($β$=12, 25 *mm*). This feature makes the proposed CCWPT system advantageous for applications where the misalignment of the plates is inevitable. The variation of $|S_{21}|$ for different rotation angles is low at center frequency (1.77 GHz) for different values of rotation angle (α). The variation is less than 1 dB.

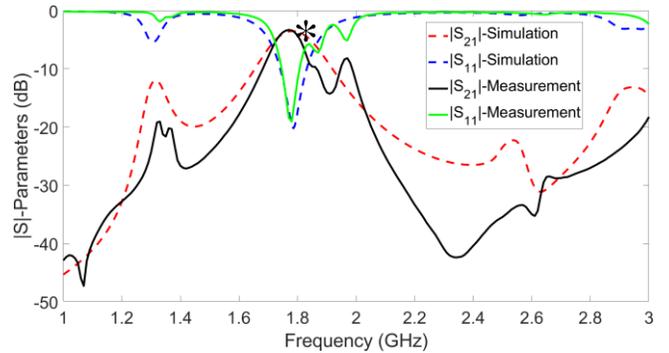

Fig. 10. Simulated and measured, results and deep learning prediction (star) of $|S_{11}|$ and $|S_{21}|$ for the fabricated prototype.

TABLE III
COMPARISON TABLE OF MEASUREMENT, SIMULATION AND THE PROPOSED AI-BASED METHOD

| Parameters | Measurement | Simulation (CST) | AI-based method |
|---|---|---|---|
| Resonance frequency (GHz) | 1.77 | 1.8 | 1.88 |
| $|S_{21}|$ (dB) | -2.5 | -2.2 | -2.32 |

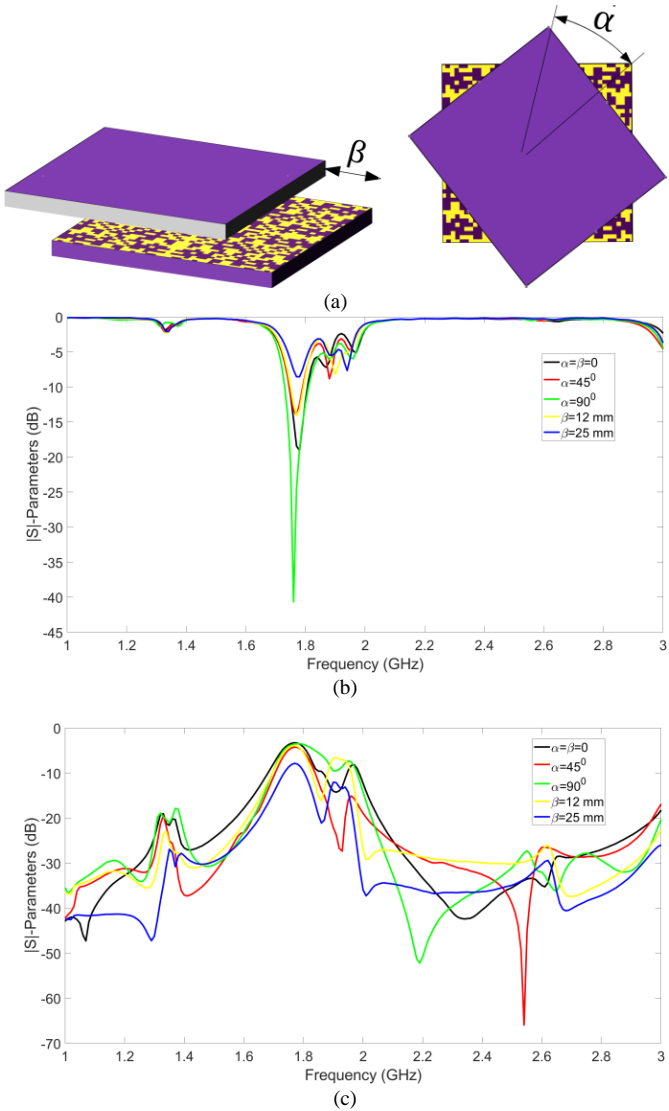

(a)

(b)

(c)

Fig. 11. Measurement results of misalignment for different misalignment parameters (α and β) a) Misaligned parallel pixelated plates, b) measured |S$_{11}$| and c) measured |S$_{21}$|.

Moreover, to illustrate the advantages of the proposed method more intuitively, we compared the AI-based iterative method and CST-based iterative method in terms of design time, number of iterations, and area between computed S-parameters and target S-parameters, as shown in Table IV. The design time of the proposed AI-based iterative method for 1000 iterations is 248 seconds on CPU and 67.9 seconds on GPU. Whereas the CST-based method takes around 250 hours for 1000 iterations. Therefore, the AI-based iterative design process on CPU is nearly 3629 times faster than the CST-based method. Therefore, the AI-based method provides an efficient and ultra-fast way of designing pixelated microstrip structures which can be used in various kinds of microwave devices (filter, coupler, antenna, metasurfaces, etc.).

*B. CCWPT System Performance*

To implement the proposed Capacitive Coupled Wireless Power Transfer (CCWPT) system, a rectifier and voltage doubler are essential components. These components are integrated into the receiver circuit, serving the purpose of converting the received RF signal into a direct current (DC) voltage, as illustrated in Fig. 2. Specifically, a Schottky diode model SMS7621 is employed within both the voltage doubler and rectifier circuits. The load resistance (RL) is set at 11 kΩ to optimize the performance. The diode's Spice parameters, crucial for accurate modeling, are detailed in Table V.

To ensure efficient power transfer to the rectifier, a matching circuit is strategically designed. This matching circuit employs two open-ended stubs, working in tandem to facilitate maximum power transfer to the rectifier component. This design choice is pivotal in enhancing the overall energy conversion efficiency of the system.

For the practical realization of the rectifier and the entire CCWPT system, physical prototypes are fabricated. Fig. 12(a) showcases the physical prototype of the rectifier, while Fig. 12(b) offers a view of the complete CCWPT system. It's important to note that the dielectric medium between the parallel plates is foam with a relative permittivity $\varepsilon_r = 1.2$, further contributing to the system's operational characteristics. According to Fig. 13, the best matching condition occurs around an input power of $-10$ *dBm*. Fig.14, shows the output DC voltage of the rectifier circuit with respect to the input power level. The output DC voltage is more than 1.5 V for an input power of -5 *dBm*.

Fig. 15 exhibits the simulated and measured RF to DC conversion efficiency of the proposed rectifier at resonance frequency of 1.77 GHz. Further, 50 % efficiency is achieved at a low power level of -5 dBm using double stub matching. According to Fig. 14 and Fig. 15, the simulation and measurement results are in good agreement, which confirms the accuracy of the design procedure.

Measurement results of the proposed CCWPT (Fig. 12(b)) are presented in Fig. 16. According to this figure, the output DC power of the CCWPT system for an input power level of 5 *dBm* is more than 1 *mW*.

TABLE IV
PERFORMANCE COMPARISON OF CST-BASED AND AI-BASED OPTIMIZATION TECHNIQUES

| Computer configuration | Method | Iteration of Computation | Design Time |
|---|---|---|---|
| Intel(R) Xeon(R) Gold 5217 CPU @ *3GHz* and 2.99 *GHz* (2 processors) /64GB/1T SSD | CST-Based Optimization Method | 1000 | 250 hours (10 days and 10 hours) |
| Intel(R) Core (TM) i7-8250U CPU @ 1.6*GHz* to 1.8 *GHz*/8GB/256G SSD | AI-Based Optimization Method on CPU | 1000 | 248 seconds |
| Tesla V100 GPU Memory (16 GB) | AI-Based Optimization Method on GPU | 1000 | 67.9 seconds |

TABLE V
SPICE PARAMETERS OF SMS7621

| $B_V(V)$ | $C_{J0}(pF)$ | $E_G(eV)$ | $I_{BV}(A)$ | $I_S(A)$ | $N$ | $V_J(V)$ | $R_S(\Omega)$ |
|---|---|---|---|---|---|---|---|
| 3 | 0.1 | 0.69 | 1×E-5 | 4×E-8 | 1.05 | 0.51 | 12 |

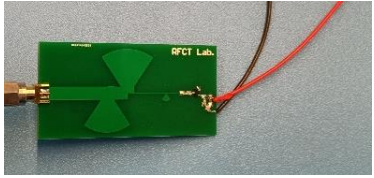
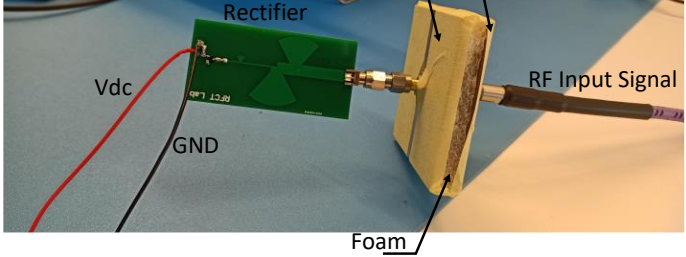

(a)

Parallel Pixelated Plates

Rectifier

Vdc

GND

RF Input Signal

Foam

(b)

Fig. 12. Fabricated, a) rectifier, b) complete CCWPT system.

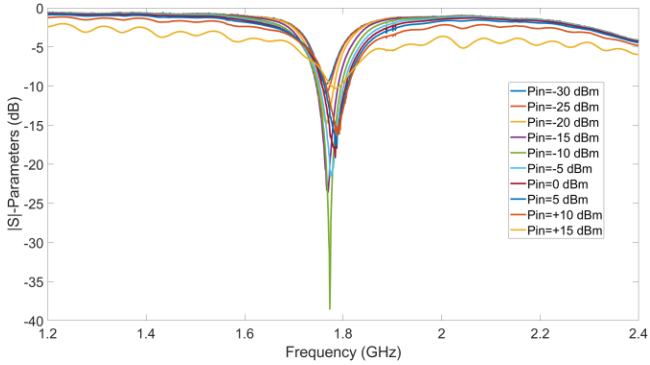

Fig. 13. Measured |S$_{11}$| at different input power levels.

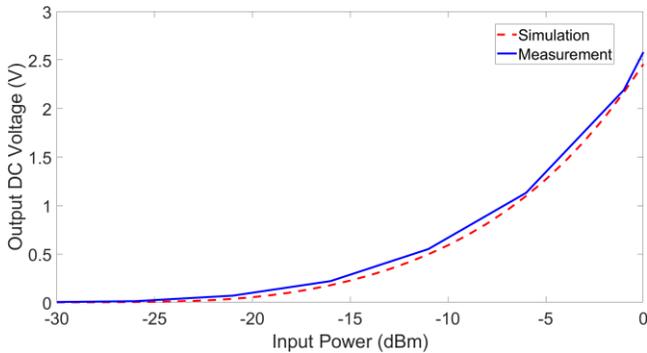

Fig. 14. Simulated and measured output DC voltage as a function of input RF power at 1.77 *GHz*.

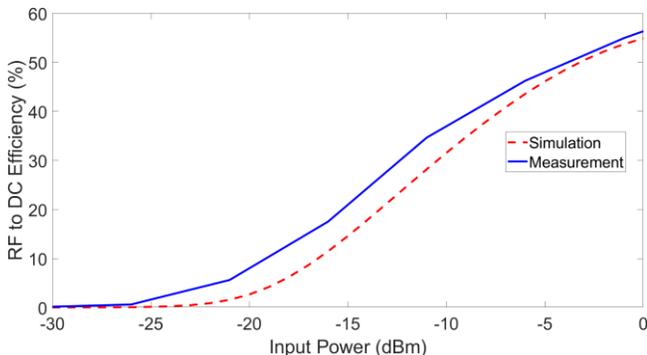

Fig. 15. Simulated and measured rectifier efficiency as a function of input RF power at 1.77 *GHz*.

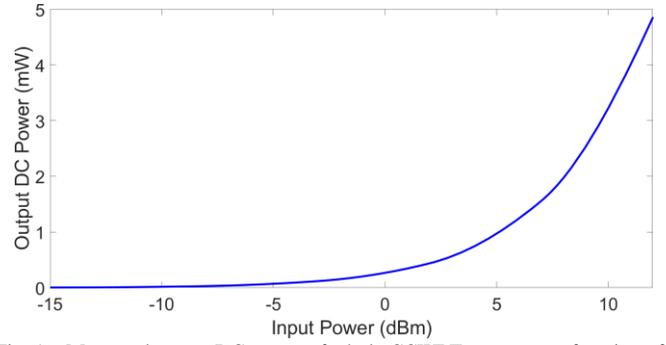

Fig. 16. Measured output DC power of whole CCWPT system as a function of input RF power at 1.77 *GHz*.

The output power of the proposed CCWPT system is 0.5 mW (500 $\mu$W) at 0 dBm input power (Fig. 16). To demonstrate the practicality of the proposed system, Table VI provides characteristics of typical sensors that have been used in agriculture and industry. According to this table, the produced DC power of 500 $\mu$W is able to turn on the sensors in Table V, instantaneously. However, 500 $\mu$W is an instantaneous power and can be collected in a battery or supercapacitor through a power management unit for a long time, and then to be consumed in a specific time. Further, recently reported WPT circuits are thoroughly compared in Table VII. According to this Table, the proposed rectifier exhibits good efficiency and sensitivity at -5 dBm in comparison with other works. This proves the usefulness of the proposed rectifier for CCWPT system.

In the fabricated structure, a mid-power rectifier is integrated at the coupler's output to demonstrate the CCWPT concept. This power range aligns well with applications such as biomedical devices and low-power IoT devices. If the intention is to develop a high-power CCWPT system, a simple adjustment would entail replacing the existing rectifier with a higher-power model. This modification facilitates effortless scalability, accommodating a diverse range of power requirements. Our emphasis on showcasing the feasibility of compact couplers and effective power transmission establishes a robust basis for future innovations and practical implementations.

TABLE VI
CHARACTERISTICS OF TYPICAL SENSORS

| Parameters | [42] | [43] | [44] | [45] | [46] |
|---|---|---|---|---|---|
| DC Supply Voltage (V) | 1 | 0.3 | 0.8 | 0.45 | 0.3 |
| Power Consumption ($\mu$W) | 150 | 10 | 15 | 14 | 3.7 |

TABLE VII
COMPARISON OF THE PROPOSED RECTIFIER WITH OTHER WORKS AT INPUT POWER OF -5 dBm

| Parameters | [47] | [48] | [49] | [50] | [51] | This Work |
|---|---|---|---|---|---|---|
| Frequency (GHz) | 5.8 | 2.4 & 5.8 | 2.4 | 5.8 | 0.4 | 1.8 |
| RF to DC efficiency (%) | 50 | 40 | 45 | 40 | 20 | 55 |

## IV. Conclusion

In this paper, an AI-based efficient and rapid design method is proposed for pixelated CCWPT microstrip structures based on deep learning neural network. This method enables the automatic and agile design procedure by iterative optimization algorithms (Genetic, BPSO, etc.) without using EM software (CST, HFSS, etc.) to calculate the S-parameter of the pixelated structure in each iteration. AI-based iterative method design process on CPU is nearly 3629 times faster than the CST-based method and shows an obvious reduction in both computational and man-powered resources. The MAE error of predicted resonance frequency and corresponding $|S_{21}|$ are 110 $MHz$ and 0.18 $dB$, respectively. To prove the concept, we designed and fabricated two parallel pixelated microstrip plates using AI-based method. The achieved consistency of the resonance frequency and $|S_{21}|$ value with the desired target demonstrates the effectiveness of the proposed method. Moreover, this method does not require any expertise in EM theory. The proposed flexible design methodology based on pixelated microstrip structure and ResNet-18 network can be used to design different microwave devices. To realize the whole CCWPT system, we designed, fabricated and tested a rectifier circuit. The measured efficiency of the rectifier is 50% at an input power of -5 dBm. Finally, we connected the rectifier to the pixelated parallel plates and measured the output DC voltage against input RF power.